# DEEP LEARNING BASED PHASE RECONSTRUCTION FOR SPEAKER SEPARATION: A TRIGONOMETRIC PERSPECTIVE

*Zhong-Qiu Wang♪, Ke Tan♪, DeLiang Wang♪,♫*

♪Department of Computer Science and Engineering, The Ohio State University, USA
♫Center for Cognitive and Brain Sciences, The Ohio State University, USA
{wangzhon, dwang}@cse.ohio-state.edu, tan.650@osu.edu

## ABSTRACT

This study investigates phase reconstruction for deep learning based monaural talker-independent speaker separation in the short-time Fourier transform (STFT) domain. The key observation is that, for a mixture of two sources, with their magnitudes accurately estimated and under a geometric constraint, the absolute phase difference between each source and the mixture can be uniquely determined; in addition, the source phases at each time-frequency (T-F) unit can be narrowed down to only two candidates. To pick the right candidate, we propose three algorithms based on iterative phase reconstruction, group delay estimation, and phase-difference sign prediction. State-of-the-art results are obtained on the publicly available wsj0-2mix and 3mix corpus.

## 1. INTRODUCTION

Audio source separation concerns the separation of a $C$-source discrete time-domain mixture $y = \sum_{c=1}^{C} s^{(c)}$ to its individual time-domain sources $s^{(c)}$. It has many applications such as speech denoising ($C = 2$, speech vs. noise) and speaker separation ($C \geq 2$, speech vs. speech). As speech is short-time stationary, a common approach is to decompose the time-domain mixture to frequency domain to reveal its frequency components using STFT and perform separation therein. One major recent advance is the introduction of deep neural networks (DNN) for the estimation of the ideal binary or ratio mask (IBM or IRM), the spectral magnitude mask (SMM) [1], or the phase-sensitive mask (PSM), where source separation is converted to a magnitude-domain T-F unit level classification or regression problem, typically retaining the mixture phase for re-synthesis. Notable works include masking based speech enhancement studies [1]–[3], and speaker separation studies such as deep clustering (DC) [4]–[6] and permutation invariant training (PIT) [7], [8]. These studies suggest that magnitude estimation can be substantially improved using deep learning based T-F masking.

In this context, this study investigates magnitude based methods for phase reconstruction for monaural speaker separation. The key insight is that the possible solutions of phase can be significantly narrowed down given sufficiently accurate magnitude estimates, under the following geometric constraint in the STFT domain:

$$Y_{t,f} = \sum_{c=1}^{C} S_{t,f}^{(c)} = \sum_{c=1}^{C} A_{t,f}^{(c)} e^{j\theta_{t,f}^{(c)}}, \quad (1)$$

where $S_{t,f}^{(c)}$ and $Y_{t,f}$ respectively denote the STFT values of source signal $c$, $s^{(c)}$, and the mixture signal $y$ at time $t$ and frequency $f$, and $A_{t,f}^{(c)} = |S_{t,f}^{(c)}|$ and $\theta_{t,f}^{(c)} = \angle S_{t,f}^{(c)}$ are the magnitude and phase of $S_{t,f}^{(c)}$, respectively. In the simplest case, suppose that there are only two sources and the two magnitude spectrums can be perfectly estimated (i.e. $\hat{A}_{t,f}^{(c)} = A_{t,f}^{(c)}$), are there any closed-form solution for phase estimation? It would be reasonable to say yes as there are two equations with two unknowns:

$$|Y_{t,f}|\cos(\angle Y_{t,f}) = \hat{A}_{t,f}^{(1)}\cos(\hat{\theta}_{t,f}^{(1)}) + \hat{A}_{t,f}^{(2)}\cos(\hat{\theta}_{t,f}^{(2)}) \quad (2)$$

$$|Y_{t,f}|\sin(\angle Y_{t,f}) = \hat{A}_{t,f}^{(1)}\sin(\hat{\theta}_{t,f}^{(1)}) + \hat{A}_{t,f}^{(2)}\sin(\hat{\theta}_{t,f}^{(2)}) \quad (3)$$

However, the underlying phase cannot be determined, because

This research was supported in part by an NSF grant (IIS-1409431), an NIDCD grant (R01 DC012048), and the Ohio Supercomputer Center.

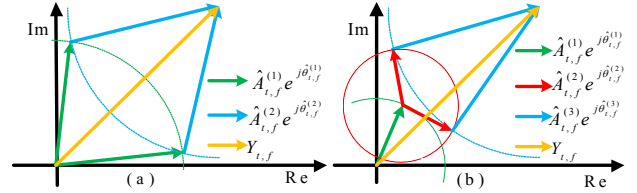

Fig. 1. Illustration of sign ambiguity when magnitudes are known in the complex plane. (a) Two-source case; (b) three-source case: for each $\hat{\theta}_{t,f}^{(1)}$, there could be two solutions for $\hat{\theta}_{t,f}^{(2)}$ and $\hat{\theta}_{t,f}^{(3)}$.

depending on the sign of the phase difference, there are two candidates satisfying the above two equations:

$$\hat{\theta}_{t,f}^{(1)} = \angle Y_{t,f} \pm \arccos((|Y_{t,f}|^2 + \hat{A}_{t,f}^{(1)^2} - \hat{A}_{t,f}^{(2)^2})/(2|Y_{t,f}|\hat{A}_{t,f}^{(1)})) \quad (4)$$

$$\hat{\theta}_{t,f}^{(2)} = \angle Y_{t,f} \mp \arccos((|Y_{t,f}|^2 + \hat{A}_{t,f}^{(2)^2} - \hat{A}_{t,f}^{(1)^2})/(2|Y_{t,f}|\hat{A}_{t,f}^{(2)})) \quad (5)$$

as is also suggested in earlier studies [9], [10]. See Fig. 1(a) for an illustration. Intuitively, this sign ambiguity occurs because the phase of each source could be either ahead of or behind the mixture phase within each T-F unit in an almost random way, posing fundamental difficulties for phase estimation. One thing we can conclude, though, is that one of the two candidates is the true $\theta_{t,f}^{(1)}$ and $\theta_{t,f}^{(2)}$.

To resolve this sign ambiguity, we think that inter-T-F unit phase relations such as group delay (GD) or instantaneous frequency [11] and phase regularizations such as phase consistency [12] could help. We propose three algorithms for phase reconstruction, leveraging good magnitude estimates produced by DNNs. The first one uses estimated magnitudes to drive an iterative phase reconstruction algorithm, which could implicitly resolve the sign ambiguity. The second one finds a sign assignment per T-F unit such that the resulting GD is closest to an estimated one. The third one implicitly predicts a sign at each T-F unit within a neural network that enforces the geometric constraint in Eq. (1).

For a mixture with $C \geq 3$, even if the magnitudes are known, there are still infinite numbers of phase candidates satisfying the geometric constraint, as is illustrated in Fig. 1(b). This suggests that it could be helpful to approach multi-source separation from a *one-vs.-the-rest* angle, where a model is trained to estimate the magnitude of source $c$ and the magnitude of the rest sources combined (denoted as $\neg c$), and at run time, the model is applied once for each source for separation. This way, there are only two possible phase candidates at each T-F unit to resolve for each source. For speaker separation, our study hence first uses a chimera++ network [5] to perform $C$-speaker separation to resolve the permutation problem and then uses an enhancement network taking into account the initial separation results of source $c$ to further estimate the magnitudes of source $c$ and $\neg c$ for phase reconstruction. Our best performing algorithm achieves state-of-the art performance on the public wsj0-2mix and 3mix dataset [4].

Why do we rely so much on magnitude estimates for phase reconstruction? This is because magnitude is much more structured and predictable than phase, and also more stable. Even if the signal is shifted slightly, the magnitude remains almost unchanged, while the phase will exhibit a phase change at every frequency and be-

**Input**: Estimated magnitudes $\hat{A}^{(c')}$ and starting phases $\hat{\vartheta}^{(c')}(0)$ initialized as mixture phase $\angle Y$ or enhanced phase $\hat{\theta}^{(c')}$ for $c'$ in $\{c, \neg c\}$, and iteration number $K$;
**Output**: Reconstructed phase $\hat{\vartheta}^{(c')}(K)$ of source $c'$, for $c'$ in $\{c, \neg c\}$;
**For** $k = 1:K$ **do**
    (1) $\hat{s}^{(c')}(k) = \text{iSTFT}(\hat{A}^{(c')}, \hat{\vartheta}^{(c')}(k-1))$, for $c'$ in $\{c, \neg c\}$;
    (2) $\varepsilon(k) = y - \sum_{c' \in \{c, \neg c\}} \hat{s}^{(c')}(k)$;
    (3) $\hat{\vartheta}^{(c')}(k) = \angle \text{STFT}(\hat{s}^{(c')}(k) + \varepsilon(k)/2)$, for $c'$ in $\{c, \neg c\}$;
**End**
Algo. 1. Two-source MISI. iSTFT($\cdot,\cdot$) reconstructs a time-domain signal from a magnitude and a phase. STFT($\cdot$) computes the magnitude and phase of a signal.

come very random if phase wrapping is incurred [11]. In addition, good magnitude estimation is achievable as is indicated in recent advance on deep learning based speech separation [1].

## 2. CHIMERA++ NETWORKS

For speaker separation, we need to first resolve the label-permutation problem. This section reviews chimera++ networks [5], which combine DC and PIT in a multi-task learning way, producing significant improvements over the individual models.

The DC algorithm [4] projects each T-F unit $i$ to a unit-length $D$-dimension vector such that the embedding matrix $V \in \mathbb{R}^{TF \times D}$, obtained by vertically stacking all the embedding vectors, can approximate the affinity matrix computed from the label matrix $U \in \mathbb{R}^{TF \times C}$, where the $i^{th}$ row is a one-hot vector denoting which of the $C$ sources dominates T-F unit $i$. Our recent study [5] suggests an alternative loss function, which whitens the embeddings in a K-means objective, leads to better performance:

$$\mathcal{L}_{DC,W} = ||V(V^TV)^{-1/2} - U(U^TU)^{-1}U^TV(V^TV)^{-1/2}||_F^2 \quad (6)$$

The PIT algorithm [7] minimizes the minimum utterance-level loss of all the permutations. The PSM [13] is typically used as the training target. The loss function is defined as:

$$\mathcal{L}_{PIT} = \min_{\pi \in \Psi} \sum_{c=1}^{C} \left\| \hat{M}^{\pi(c)} \otimes |Y| - T_0^{|Y|}(|S^{(c)}| \otimes \cos(\angle S^{(c)} - \angle Y)) \right\|_1, \quad (7)$$

where $\Psi$ is a set of permutations over $C$ sources, $\hat{M}$ is the estimated PSM, and $\otimes$ denotes element-wise multiplication. Using $T_a^b(\cdot) = \max(a, \min(b, \cdot))$, $T_0^{|Y|}(\cdot)$ truncates the PSM to the range [0,1]. Sigmoidal units are used as the output non-linearity to obtain $\hat{M}^{(c)}$.

The loss function of chimera++ networks is a weighted combination of the above two loss functions:

$$\mathcal{L}_{chi++} = \lambda \mathcal{L}_{DC,W} + (1-\lambda)\mathcal{L}_{PIT} \quad (8)$$

At run time, the PIT output is utilized for separation.

In [5], a vanilla bi-directional LSTM (BLSTM) is used in the chimera++ network. To improve mask estimation, we employ a BLSTM with convolutional encoder-decoder structures and skip connections [14], which will be discussed in Section 4.

## 3. PROPOSED ALGORITHMS

With the label-permutation problem resolved, an enhancement network, which includes the estimated mask $\hat{M}^{(c)}$ produced by the chimera++ network as inputs, is trained for each of the following three proposed algorithms to further estimate the magnitude of source $c$ and $\neg c$ for phase reconstruction. See Fig. 2 for the network architectures. A side product of this research is a new way of computing the PSM using magnitude estimates (see Section 3.4).

### 3.1. Deep Learning Based Iterative Phase Reconstruction

One straightforward approach for phase reconstruction is to use estimated magnitudes to drive an iterative phase reconstruction algorithm [15], [16], [5], [6]. Here, we employ the multiple input spectrogram inverse (MISI) algorithm [17] (see Algo. 1). Our insight is that the error distribution step (see (2) and (3) in Algo. 1) can ensure that the estimated phases are taken from reconstructed signals that add up to the mixture signal. The geometric constraint is hence roughly satisfied. If the magnitudes of the reconstructed signals are sufficiently accurate, the signs of many T-F units could be automatically determined, because the reconstructed signals are real signals that guarantee to have consistent phase structures and only particular ways of sign assignments exhibit consistent phase.

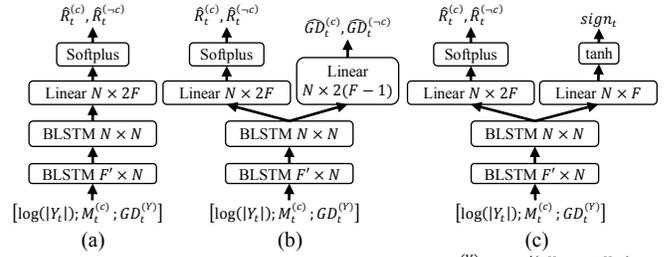

Fig. 2. Enhancement network architectures. Note that $GD_{t,f}^{(Y)} = \angle e^{j(\angle Y_{t,f+1} - \angle Y_{t,f})}$.

One issue with our recent studies [5], [6] employing MISI for phase reconstruction is that the PSM is used as the training target in PIT and the resulting magnitude estimates are used for MISI. However, the sum of such magnitude estimates almost equals the mixture magnitude, as the sum of the PSMs of all the sources is one. Under the geometric constraint, the most reasonable phase estimate for each source is therefore simply the mixture phase. For example, in Fig. 1(a), if $\hat{A}_{t,f}^{(1)} + \hat{A}_{t,f}^{(2)} = |Y_{t,f}|$, the three sides cannot make a triangle and the absolute phase difference estimates $|\hat{\theta}_{t,f}^{(1)} - \angle Y_{t,f}|$ and $|\hat{\theta}_{t,f}^{(2)} - \angle Y_{t,f}|$ are both zero. Similar issues will be incurred if the sum of estimated magnitudes is implicitly or explicitly constrained to equal the mixture magnitude, such as using the IBM or IRM as the training target, using softmax as the output non-linearity, and estimating noise magnitude by subtracting estimated speech magnitude from the mixture magnitude.

This study hence estimates the SMM by using the loss function in Eq. (9), rather than the PSM using Eq. (10). See Fig. 2(a) for the network structure. This minor change leads to large improvements in our experiments after MISI is applied for phase reconstruction.

$$\mathcal{L}_{MSA(\alpha)}^{Enh1} = \mathcal{L}_{MSA(\alpha)} = \sum_{c' \in \{c, \neg c\}} \left\| |Y| \otimes T_0^{\alpha}(\hat{R}^{(c')}) - T_0^{\alpha|Y|}(|S^{(c')}|) \right\|_1 \quad (9)$$

where $\hat{R}^{(c')}$ is the estimated SMM obtained by using Softplus non-linearity. Based on the trigonometric perspective, $\alpha$ should be much larger than one so that the estimated magnitudes can be large enough compared with the mixture magnitude when necessary to elicit a large enough phase difference for phase reconstruction, such as when the sources cancel with each other at a T-F unit.

To facilitate comparison, we also train the same network with minimal changes to estimate the PSM using the following loss:

$$\mathcal{L}_{PSA(\gamma,\beta)}^{Enh1} = \sum_{c' \in \{c, \neg c\}} \left\| |Y| \otimes T_{\gamma}^{\beta}(\hat{Q}^{(c')}) - T_{\gamma|Y|}^{\beta|Y|}(|S^{(c')}| \otimes \cos(\angle S^{(c')} - \angle Y)) \right\|_1 \quad (10)$$

where the estimated PSM $\hat{Q}^{(c')}$ is obtained by using sigmoid activation when $\beta=1$ and $\gamma=0$, linear activation when $\beta>1$ and $\gamma<-1$, Softplus when $\beta>1$ and $\gamma=0$, and tanh when $\beta=1$ and $\gamma=-1$.

Following [6], we unfold the MISI iterations as multiple layers in the network and compute the loss function in the time domain.

$$\mathcal{L}_{MISI-K}^{Enh1} = \sum_{c' \in \{c, \neg c\}} \left\| \text{iSTFT}(\hat{A}^{(c')}, \hat{\vartheta}^{(c')}(K)) - s^{(c')} \right\|_1, \quad (11)$$

where $\hat{\vartheta}^{(c')}(K)$ denotes the reconstructed phase after $K$ iterations of MISI (see Algo. 1 for detailed definitions), which starts from estimated magnitude $\hat{A}^{(c')} = |Y| \otimes \hat{R}^{(c')}$ and the mixture phase $\angle Y$.

### 3.2. Group Delay Based Phase Reconstruction

For a pair of T-F units at two consecutive frequencies, there are four ($2^2$) combinations of possible phase solutions, while only one combination exhibits a particular group delay. Our study first estimates the group delay of each source and then finds a sign assignment at each T-F unit in a way such that the resulting phase spectrum has a group delay closest to the estimated one. Note that group delay (GD) [18], computed as $GD_{t,f}^{(c)} = \angle e^{j(\angle S_{t,f+1}^{(c)} - \angle S_{t,f}^{(c)})}$, exhibits patterns clearly predictable from (see Fig. 3), and is mathematically related to, log magnitude [11], [19]–[21].

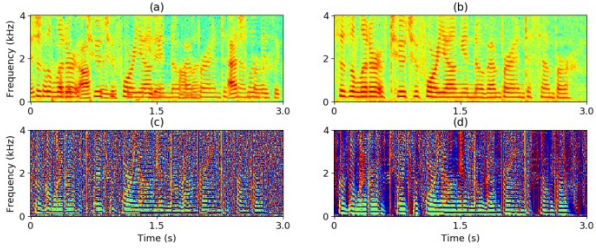

Fig. 3. Illustration of GD using a two-speaker mixture. (a) Log magnitude of mixture; (b) log magnitude of source 1; (c) clean GD of source 1; (d) estimated GD of source 1.

The network structure is depicted in Fig. 2(b). The loss function for the GD branch is:

$$\mathcal{L}_{GD1} = \sum_{c'\in\{c,\neg c\}}\sum_{t}\sum_{f=1}^{F-1} |S_{t,f+1}^{(c')}|(1-\cos(\widehat{GD}_{t,f}^{(c')} - GD_{t,f}^{(c')}))/2, \quad (12)$$

and the overall loss function is: $\mathcal{L}_{MSA(\alpha)+GD1}^{Enh2} = \mathcal{L}_{MSA(\alpha)} + \mathcal{L}_{GD1}$.

At run time, assuming that $\hat{A}^{(c)}$, $\hat{A}^{(\neg c)}$ and $|Y|$ form a triangle at each T-F unit, we first estimate the absolute phase difference $\hat{\delta}^{(c')}$ between source $c'$ and the mixture based on the law of cosines:

$$\hat{\delta}^{(c')} = |\angle e^{j(\hat{\theta}^{(c')} - \angle Y)}| = \arccos(\mathcal{T}(\frac{|Y|^2 + \hat{A}^{(c')^2} - \hat{A}^{(\neg c')^2}}{2|Y|\otimes|\hat{A}^{(c')}|})), \text{ for } c' \text{ in } \{c,\neg c\} \quad (13)$$

where $\mathcal{T}(\cdot)$ truncates the values outside of the range $[-1,1]$ to 1. Note that when $\hat{A}_{t,f}^{(c)} + \hat{A}_{t,f}^{(\neg c)} \leq |Y_{t,f}|$, the three sides cannot make a triangle. This can happen as we are using estimated magnitudes. In addition, $\hat{A}^{(c)}$ and $\hat{A}^{(\neg c)}$ could have zero values in some T-F units, if obtained via ReLU. We hence clip the values outside the range $[-1,1]$ to 1, meaning that the mixture phase is considered as the phase estimate for such T-F units since $\arccos(1) = 0$.

We then determine the sign assignment at each T-F unit, $\hat{g}_{t,f} \in \{-1,1\}$, by maximizing the following objective at each frame:

$$\hat{g}_{t,1},\ldots,\hat{g}_{t,F} = \underset{g_{t,1},\ldots,g_{t,F}}{\mathrm{argmax}} \sum_{f=1}^{F-1}\sum_{c'\in\{c,\neg c\}} \cos\left(\hat{\theta}_{t,f+1}^{(c')}(g_{t,f+1}) - \hat{\theta}_{t,f}^{(c')}(g_{t,f}) - \widehat{GD}_{t,f}^{(c')}\right), \quad (14)$$

where $\hat{\theta}_{t,f}^{(c)}(g_{t,f})$ and $\hat{\theta}_{t,f}^{(\neg c)}(g_{t,f})$ are phases hypothesized as:

$$\hat{\theta}_{t,f}^{(c)}(g_{t,f}) = \angle Y_{t,f} + g_{t,f}\hat{\delta}_{t,f}^{(c)} \quad (15)$$
$$\hat{\theta}_{t,f}^{(\neg c)}(g_{t,f}) = \angle Y_{t,f} - g_{t,f}\hat{\delta}_{t,f}^{(\neg c)} \quad (16)$$

Although Eq. (14) has $2^F$ possible solutions, our insight is that it can be efficiently solved with time complexity $O(2^2 F)$ by applying dynamic programming (or Viterbi decoding) within each frame, as the estimated GD only characterizes the phase relations between each consecutive T-F unit pair along frequency. The final phase estimates are obtained as $\angle Y + \hat{g}\otimes\hat{\delta}^{(c)}$ and $\angle Y - \hat{g}\otimes\hat{\delta}^{(\neg c)}$.

There are previous studies [9], [10] employing GD for sign determination. However, they resolve the ambiguity using an empirically hypothesized minimum GD deviation constraint and only consider a few frequencies with detected harmonic peaks.

### 3.3. Sign Prediction Networks

The GD based method is designed to be applied at run time as post processing. It is hard to perform end-to-end training. A possibly better approach is to let the network predict the sign explicitly (see Fig. 2(c)), and compute the estimated phases as follows:

$$\hat{\theta}^{(c)} = \angle Y + sign\otimes\hat{\delta}^{(c)} \quad (17)$$
$$\hat{\theta}^{(\neg c)} = \angle Y - sign\otimes\hat{\delta}^{(\neg c)} \quad (18)$$

where $sign$ is obtained via tanh non-linearity. Note that $\hat{\delta}^{(c')}$ is naturally bounded in the range $[0,\pi]$ and $sign\otimes\hat{\delta}^{(c')}$ in the range $[-\pi,\pi]$. The loss function on estimated phases is:

$$\mathcal{L}_{GD2} = \sum_{c'\in\{c,\neg c\}}\sum_{t}\sum_{f=1}^{F-1}|S_{t,f+1}^{(c')}|(1-\cos(\hat{\theta}_{t,f+1}^{(c')} - \hat{\theta}_{t,f}^{(c')} - GD_{t,f}^{(c')}))/2, \quad (19)$$

and the overall loss function is: $\mathcal{L}_{MSA(\alpha)+GD2}^{Enh3} = \mathcal{L}_{MSA(\alpha)} + \mathcal{L}_{GD2}$. This way, the network could learn to produce a sign that can lead to GD spectrums close to the clean ones. An alternative is to compute the loss from the estimated phases and clean phases:

$$\mathcal{L}_{phase} = \sum_{c'\in\{c,\neg c\}} \left\||S^{(c')}|\otimes(1-\cos(\hat{\theta}^{(c')} - \theta^{(c')}))/2\right\|_1, \quad (20)$$

and the overall loss function is: $\mathcal{L}_{MSA(\alpha)+phase}^{Enh3} = \mathcal{L}_{MSA(\alpha)} + \mathcal{L}_{phase}$.

We emphasize that Eq. (17) and (18) (as well as (15) and (16)) implicitly constrain that, at each T-F unit, the two reconstructed STFT vectors ($\hat{A}_{t,f}^{(c)}e^{j\hat{\theta}_{t,f}^{(c)}}$ and $\hat{A}_{t,f}^{(\neg c)}e^{j\hat{\theta}_{t,f}^{(\neg c)}}$) have to be on the different sides of the mixture STFT vector $Y_{t,f}$ in the complex plane, and $\hat{\theta}_{t,f}^{(c)}$ and $\hat{\theta}_{t,f}^{(\neg c)}$ cannot be, at the same time, more than $\pi/2$ away from $\angle Y_{t,f}$, because only in this way could the two reconstructed STFT vectors add up to the mixture STFT vector. This distinguishes our approach from studies that directly predict unbounded or unconstrained phase differences [22], [23], clean phases [24] and real and imaginary components of target sources [25], [26], or fully complex neural network approaches [27].

Following our recent study [6], we train through iSTFT for time-domain waveform approximation (WA), using $\hat{A}^{(c')}$ and $\hat{\theta}^{(c')}$:

$$\mathcal{L}_{WA}^{Enh3} = \sum_{c'\in\{c,\neg c\}} \left\|\text{iSTFT}(\hat{A}^{(c')}, \hat{\theta}^{(c')}) - s^{(c')}\right\|_1 \quad (21)$$

Following [23], which uses estimated phases as the starting phases to train through MISI, we further train our model using:

$$\mathcal{L}_{MISI-K}^{Enh3} = \sum_{c'\in\{c,\neg c\}} \left\|\text{iSTFT}(\hat{A}^{(c')}, \hat{\vartheta}^{(c')}(K)) - s^{(c')}\right\|_1, \quad (22)$$

where $\hat{\vartheta}^{(c')}(K)$ is obtained after $K$ iterations of MISI starting from $\hat{A}^{(c')}$ and $\hat{\theta}^{(c')}$ produced by the sign prediction network. We will denote $\mathcal{L}_{WA}^{Enh3}$ as $\mathcal{L}_{MISI-0}^{Enh3}$, since $\hat{\vartheta}^{(c')}(0) = \hat{\theta}^{(c')}$ (see Algo. 1).

Following [6], [28], [29], which computes loss using the magnitudes of reconstructed signals, we further train the network using:

$$\mathcal{L}_{MISI-K-MSA}^{Enh3} = \sum_{c'\in\{c,\neg c\}}\left\|\left|\text{STFT}\left(\text{iSTFT}\left(\hat{A}^{(c')}, \hat{\vartheta}^{(c')}(K)\right)\right)\right| - |S^{(c')}|\right\|_1 \quad (23)$$

Our insight is that due to phase inconsistency, the reconstructed signal, $\text{iSTFT}(\hat{A}^{(c')}, \hat{\vartheta}^{(c')}(K))$, may not exhibit a magnitude as good as $\hat{A}^{(c')}$, although the iterative process in MISI can reduce their difference [30]. The network trained this way outputs two signals that almost add up to the mixture signal and each signal is expected to exhibit a good magnitude. From the trigonometric perspective, the signs could be automatically determined because the two signals are real signals having consistent phase structures, as is explained in the first paragraph of Section 3.1.

### 3.4. Computing PSM from Estimated Magnitudes

A side product of this research is a new way of computing the PSM (defined as $|S^{(c)}|\otimes\cos(\angle S^{(c)} - \angle Y)/|Y|$) [13] in two-source cases, where the cosine term can be estimated as $\cos(\hat{\delta}^{(c)})$:

$$\hat{Z}^{(c)} = \hat{A}^{(c)}\otimes\cos(\hat{\delta}^{(c)})/|Y| \quad (24)$$

In the literature, the PSM is typically clipped to the range $[0,1]$ and directly predicted by a DNN in a way similar to Eq. (7) or using $\mathcal{L}_{PSA(0,1)}^{Enh1}$ (i.e. $\beta=1$ and $\gamma=0$) in Eq. (10) [13]. In contrast, the estimated PSM obtained here is assembled based on estimated magnitudes. It is not limited to the range $[0,1]$ and can even go negative.

## 4. EXPERIMENTAL SETUP

We validate our algorithms on the wsj0-2mix and 3mix dataset [4], designed for a talker-independent speaker separation task. Each of them contains 20,000, 5,000, and 3,000 2(or 3)-speaker mixtures in its 30, 10 and 5 h training, validation, and test (open speaker condition, OSC) set, respectively. The sampling rate is 8 kHz. The SNR in each mixture is randomly sampled from -5 to 5 dB. We use 32 ms window size and 8 ms hop size. Square-root Hann window is applied before 256-point DFT is applied to extract 129-dimensional log magnitude features. $\lambda$ in Eq. (8) is set to 0.975 and embedding dimension $D$ set to 20. $K$ in MISI is set to 5.

We use a 4-layer BLSTM with convolutional encoder-decoder structures and skip connections [31], [32] in the chimera++ network (see Fig. 4). Similar network was found useful in a speech enhancement study [14]. The encoder contains 7 convolutional blocks, each including 2D convolution, batch normalization and exponential linear units (ELU). The decoder contains 6 deconvolu-tional blocks, each consisting of 2D deconvolution, BN and ELU layers, and one 2D deconvolution layer and a sigmoidal layer to obtain estimated masks. The embedding layer grows out from the last BLSTM layer. Each BLSTM has 512 units in each direction.

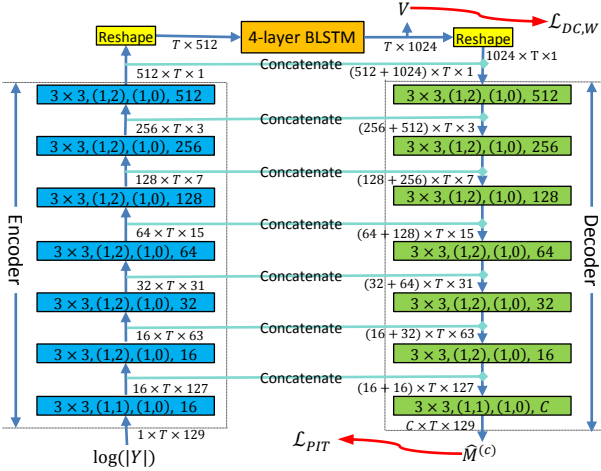

Fig. 4. Chimera++ network architecture. The tensor shape after each block is in format: *featureMaps × timeSteps × frequencyChannels*. Each block is specified in format: *kernelSizeTime×kernelSizeFreq, (stridesTime, stridesFreq), (paddingsTime, paddingsFreq), featureMaps*.

Each enhancement network (see Fig. 2) contains three BLSTM layers, each with 600 units in each direction.

We use scale-invariant signal-to-distortion ratio improvement (SI-SDRi) [33] as the major evaluation metric. We also report SDR improvement (SDRi) computed using the *bss_eval* software [34], and perceptual estimation of speech quality (PESQ) scores [35].

## 5. EVALUATION RESULTS

In Table 1, we report the performance on wsj0-2mix. Including the encoder-decoder structure into the chimera++ network improves SI-SDRi by 0.7 dB (from 11.2 to 11.9 dB), compared with [5] that uses a vanilla BLSTM. The enhancement network, which can also be thought of as stacking [36], [37], improves estimated PSM results from 11.9 to 12.1 dB, by using $\mathcal{L}_{PSA(0,1)}^{Enh1}$. Further applying 5 iterations of MISI (MISI-5) at run time only leads to slight improvement (from 12.1 to 12.5 dB). Similar trend is observed for models trained using $\mathcal{L}_{PSA(0,5)}^{Enh1}$, $\mathcal{L}_{PSA(-1,1)}^{Enh1}$, and $\mathcal{L}_{PSA(-5,5)}^{Enh1}$. In contrast, the model trained to estimate the SMM using $\mathcal{L}_{MSA(5)}^{Enh1}$ (i.e. $\alpha$=5) exhibits substantial improvements when combined with MISI-5 (from 11.1 to 14.4 dB), indicating that the SMM is the preferred training target if MISI needs to be performed. Further training the model with $\mathcal{L}_{MISI-5}^{Enh1}$ pushes the performance to 15.0 dB. Compared with $\mathcal{L}_{MSA(5)}^{Enh1}$, using estimated group delay from $\mathcal{L}_{MSA(5)+GD1}^{Enh2}$ for phase reconstruction improves the performance from 11.1 to 13.6 dB, while this approach is not as good as the sign prediction networks that can be trained end-to-end. Compared with $\mathcal{L}_{MSA(5)}^{Enh1}$, $\mathcal{L}_{MSA(5)+phase}^{Enh3}$ and $\mathcal{L}_{MSA(5)+GD2}^{Enh3}$ both lead to substantial improvement (14.4 and 14.2 vs. 11.1 dB). The former is slightly better, likely because it directly compares estimated phases with clean ones for loss computation. Further applying MISI-5 on the estimated magnitudes and enhanced phase improves the results to 15.0 dB, which is 0.6 dB (15.0 vs. 14.4 dB) better than applying MISI-5 on the model trained with $\mathcal{L}_{MSA(5)}^{Enh1}$, indicating the benefits of using an enhanced phase as the starting phase for MISI over using the mixture phase. Further training through MISI using $\mathcal{L}_{MISI-5}^{Enh3}$ produces slight improvement (from 15.0 to 15.3 dB). Compared with $\mathcal{L}_{MISI-5}^{Enh3}$, $\mathcal{L}_{MISI-5-MSA}^{Enh3}$ leads to worse SI-SDRi (15.2 vs. 15.3 dB), which aligns with the findings in [6]. Different from [6], the PESQ score is improved significantly from 3.36 to 3.45. This could be that PESQ is computed by reducing the phase mismatch between enhanced signals and reference signals via a time alignment procedure, and considerably taking into account the magnitudes of re-synthesized signals [35], while SI-SDR solely considers time-domain signals and is hence more sensitive to phase mismatches. For the side product in Eq. (24), which assembles an estimated PSM from the estimated magnitudes produced via $\mathcal{L}_{MSA(5)}^{Enh1}$, it obtains results comparable to $\mathcal{L}_{PSA(-5,5)}^{Enh1}$, and better than the other three models trained to directly estimate the PSM.

Table 1. Average SI-SDRi (dB) and PESQ results on OSC of wsj0-2mix.

| Approaches | Models | Enhanced Phase? | SI-SDRi | PESQ |
|---|---|---|---|---|
| Unprocessed | - | No | 0.0 | 2.01 |
| Chimera++(Encoder-BLSTM-Decoder) | $\mathcal{L}_{chi++}$ | No | 11.9 | 3.12 |
| Deep learning based iterative phase reconstruction | $\mathcal{L}_{PSA(0,1)}^{Enh1}$ | No | 12.1 | 3.15 |
| | +MISI-5 | Yes | 12.5 | 3.17 |
| | $\mathcal{L}_{PSA(0,5)}^{Enh1}$ | No | 12.4 | 3.17 |
| | +MISI-5 | Yes | 12.9 | 3.19 |
| | $\mathcal{L}_{PSA(-1,1)}^{Enh1}$ | No | 12.4 | 3.21 |
| | +MISI-5 | Yes | 12.9 | 3.24 |
| | $\mathcal{L}_{PSA(-5,5)}^{Enh1}$ | No | 12.7 | 3.21 |
| | +MISI-5 | Yes | 13.3 | 3.24 |
| | $\mathcal{L}_{MSA(5)}^{Enh1}$ | No | 11.1 | 3.27 |
| | +MISI-5 | Yes | 14.4 | 3.43 |
| | +$\mathcal{L}_{MISI-5}^{Enh1}$ | Yes | 15.0 | 3.38 |
| | +Eq. (24) | No | 12.6 | 3.24 |
| Group delay based phase reconstruction | $\mathcal{L}_{MSA(5)+GD1}^{Enh2}$ | Yes | 13.6 | 3.39 |
| Sign prediction network | $\mathcal{L}_{MSA(5)+GD2}^{Enh3}$ | Yes | 14.2 | 3.39 |
| | $\mathcal{L}_{MSA(5)+phase}^{Enh3}$ | Yes | 14.4 | 3.38 |
| | +MISI-5 | Yes | 15.0 | 3.44 |
| | +$\mathcal{L}_{WA}^{Enh3}$ | Yes | 14.6 | 3.36 |
| | +$\mathcal{L}_{MISI-5}^{Enh3}$ | Yes | 15.3 | 3.36 |
| | +$\mathcal{L}_{MISI-5-MSA}^{Enh3}$ | Yes | 15.2 | 3.45 |

Table 2. Average SI-SDRi (dB), SDRi (dB) and PESQ comparison between proposed algorithms and other methods on OSC of wsj0-2mix and wsj0-3mix.

| Approaches | wsj0-2mix | | | wsj0-3mix | | |
|---|---|---|---|---|---|---|
| | SI-SDRi | SDRi | PESQ | SI-SDRi | SDRi | PESQ |
| Unprocessed | 0.0 | 0.0 | 2.01 | 0.0 | 0.0 | 1.66 |
| DC++ [4], [39] | 10.8 | - | - | 7.1 | - | - |
| ADANet [33] | 10.4 | 10.8 | 2.82 | 9.1 | 9.4 | 2.16 |
| uPIT-ST [7], [37] | - | 10.0 | - | - | 7.7 | - |
| Chimera++ (BLSTM) [5] | 11.2 | 11.5 | - | - | - | - |
| +MISI-5 [5] | 11.5 | 11.8 | - | - | - | - |
| +WA-MISI-5 [6] | 12.6 | 12.9 | - | - | - | - |
| + PhaseBook [23] | 12.8 | - | - | - | - | - |
| conv-TasNet-gLN [40], [38] | 14.6 | 15.0 | 3.25 | 11.6 | 12.0 | 2.50 |
| Proposed (sign prediction net, $\mathcal{L}_{MISI-5}^{Enh3}$) | 15.3 | 15.6 | 3.36 | 12.1 | 12.5 | 2.64 |
| Proposed (sign prediction net, $\mathcal{L}_{MISI-5-MSA}^{Enh3}$) | 15.2 | 15.4 | **3.45** | 12.0 | 12.3 | **2.77** |

In Table 2, we compare the performance of our algorithm with other competitive systems on the wsj0-2mix and 3mix corpus. Our algorithm obtains dramatically better performance than the other STFT based approaches. Its performance is also better than a recent time-domain approach [38], particularly in terms of PESQ, even though we largely rely on magnitude information.

## 6. CONCLUDING REMARKS

Thanks to a novel trigonometric perspective, we have proposed three phase reconstruction algorithms based on magnitude estimation. The obtained state-of-the-art speaker separation results suggest that deep learning based magnitude estimation can clearly benefit phase reconstruction. Future research will employ more powerful neural networks for magnitude estimation and explore this trigonometric insight for time- and complex-domain speaker separation, speech de-noising and speech de-reverberation. In closing, we emphasize that a geometric constraint affords a mechanism to narrow down the possible solutions of phase, and it could play a fundamental role in future research on phase estimation.